# Expression proteomics reveals protein targets and highlights mechanisms of action of small molecule drugs

*Alexey Chernobrovkin,[1] Consuelo Marin Vicente,[1,2] Neus Visa[2]*

*and Roman A. Zubarev[1] \**

[1]Division of Physiological Chemistry I, Department of Medical Biochemistry and Biophysics, Karolinska Institutet, Scheelesväg 2, SE-17 177 Stockholm, Sweden

[2]Department of Molecular Biology, Stockholm University, Stockholm, Sweden

**\*Corresponding author:** Roman.Zubarev@ki.se

*Running title*: Drug target by proteomics

# ABSTRACT

Phenomenological screening of small molecule libraries for anticancer activity yields potentially interesting candidate molecules, with a bottleneck in the determination of drug targets and the mechanism of anticancer action. A novel approach to drug target deconvolution compares the abundance profiles of proteins expressed in a panel of cells treated with different drugs, and identifies proteins with cell-type independent and drug-specific regulation that is exceptionally strong in relation to the other proteins. Mapping top candidates on known protein networks reveals the mechanism of drug action, while abundant proteins provide a signature of cellular death/survival pathways. The above approach can significantly shorten drug target identification, and thus facilitate the emergence of novel anticancer treatments.

Target-directed discovery is the way pharmaceutical industry most often uses in searching for new drugs, with compound libraries screened for binding or activity against a known protein target. In contrast, phenomenological screening of small molecule libraries is a "black-box" target-agnostic approach, where compounds are interrogated in cell-based assays with a readout linked to a disease-relevant process (e.g. cancer cell apoptosis). Arguably, this latter approach to drug discovery offers better chances for success because more targets are addressed and the assay is more relevant to human physiology. Indeed, between 1999 and 2008, of the first-in-class compounds that were approved by the FDA, only 38% had been derived from target-based screening, while the rest - from phenotypic screening.[1] However efficient, phenomenological screening has a serious bottleneck in drug target discovery and validation. Less than 200 small-molecule anticancer drugs approved by FDA have a known mechanism of action, while thousands of promising molecules remain with poorly known or unknown targets.[2,3] This mismatch between the number of promising compounds and the knowledge of the targets and underlying mechanisms of action represents one of the greatest unmet needs in war against cancer.

Mass-spectrometry based proteomics is a well established tool in drug discovery.[4] Unlike transcriptomics that covers the whole range of expressed genes, a typical untargeted 1D LC-MS/MS proteomics experiment can detect and quantify up to 5,000 proteins, which is less than half of the expressed human proteome.[5] However, proteomics technology rapidly progresses, and deep proteome analysis with more than 8,000 protein groups quantified is becoming increasingly available.[6–9] Besides, measuring relative protein concentrations accounts not only for protein expression but also for protein degradation, which makes proteomics particularly valuable in drug target discovery. Indeed, drug attachment often stabilizes the protein and makes it more resistant to degradation, which leads to protein target accummulation.[10,11] Alternatively, drug-induced protein translocation from nucleus to cytoplasm can cause its rapid degradation.[12]

Recently, significant hope has been associated with dynamic proteomics, which had a promising start in drug target discovery. In a landmark experiment (performed without the use of mass spectrometry), the known target for camptothecin, DNA topoisomerase 1 (TOP1), has been found among the proteins with fast and deep abundance reduction upon drug application.[12] However, detailed analysis of the dynamic proteomics data using time-dependent abundance changes as sole criteria could only identify TOP1 as the 35$^{th}$ most likely candidate among ca. 900 quantified proteins.[13]

This result epitomizes the main weakness of the dynamic proteomics approach to drug target discovery: while a multitude of proteins change their abundance upon drug application (in the cited experiment, *all* measured protein abundances have changed within 48 h),[12] many changes have low specificity in respect to the drug. Therefore, appropriate filtering ("*in silico* target purification") is needed to disregard these unspecific proteins and identify the true drug target. Pathway analysis can provide such a filter; being applied to the dynamic proteomics dataset, it identified TOP1 together with only eight other likely candidates.[13] However, while of clearly significant potential for drug discovery, the pathway analysis approach has a number of shortcomings limiting its generality. Importantly, it searches targets among the known key nodes (bottleneck regulatory molecules) in the pathway database, relying on already known information for data filtering. In essence, only known potential drug targets will be discovered, with the discovery process being database-sensitive.

An alternative approach could be to focus on early times after drug applications. The rational is that the initial cellular response should be most drug-specific, while the late response (e.g., apoptosis) is largely generic. However, in an analysis[14] of early response times of RKO cells to a broadly used anticancer substance, 5-fluorouracil (5-FU), which has a well-known target, thymidylate synthase (TYMS), the changes in the first few hours past drug treatment have been too small to be measured with sufficient statistical confidence.

Recently, we have found that the apoptotic response is much less generic than has previously been thought, and that in fact proteome changes become more apparent and drug-specific in late apoptosis. Based on this finding, we have developed a new method of drug target identification that does not require *a priori* knowledge of signaling or metabolic pathways and has a high robustness. The method achieves *in silico* target purification using an expression proteomics dataset obtained from a treatment of a panel of cell lines (≥2 cell lines) with a panel of drugs (≥3 molecules). The approach is based on the assumptions that the drug target significantly changes its abundance (up or down) in late apoptosis, and that the drug target abundance behaves similarly for different cell lines that are sensitive to the drug. It is understood that the validity of these assumptions will be a factor limiting the applicability of the approach; however, most systems tested so far complied with these assumptions. The method also capitalizes on recent developments in label-free proteomics[5,15,16] that made possible analysis of ≥5000 proteins in a reasonably short time (3-4 h). This "high content proteomics"[5] enabled in turn quantitative cross-comparison of several cell lines treated with multiple drugs, with several different times of treatments, all with adequate controls and replicate analyses. Below we describe a set of proof-of-principle experiments, where known protein targets for five drugs were correctly identified, often as top candidates, with novel plausible target candidates revealed.

**RESULTS**

*Experiment I*. Three cell lines (melanoma A375, lung cancer H1299 and colon cancer HCT116) were treated with 5-FU, methotrexate (MTX), paclitaxel (PCTL), doxorubicin (DOXO) and tomudex (TDX). In total, 5037 proteins were identified with FDR<1%. Label-free quantification across all samples was performed for 4168 proteins that were identified with at least two unique peptides.

The method's workflow is depicted in Figure 1. After LC-MS/MS based proteomics analysis of the panel of cell lines treated by a panel of drugs and controls, regulation **Reg** and specificity **Spec** were calculated for each protein, cell line and treatment. Regulation and specificity values were then subjected to rank product analysis that calculated the final ranks and p-values (using the Bonferroni correction), thus identifying significantly regulated proteins. The protein list was then sorted by p-values in ascending order. The few top proteins with statistically significant p-values represented the most likely drug targets, while all statistically significant proteins were mapped on protein-protein interaction networks to identify the mechanism of drug action. Finally, the most abundant part of the proteomes was mapped on a 2D (or higher dimensionality) plot using principal component analysis to reveal the pathways of cellular death and survival. Below we discuss in detail the implementation of the method for each of the three levels of description (drug target, mechanism of action and death/survival mode).

**Drug target**. Figure 2a) illustrates the *in silico* drug target purification procedure by showing the regulation of TUBB2A (ß-tubulin) in different cell lines and under different treatments. For all three cell lines, this protein demonstrates a consistent upregulation in PCTL treatment, which significantly exceeds the regulation levels in any other treatment or control. While the rank of TUBB2A for regulation and specificity is not higher than 7 among 4168 proteins for any cell line, the combined rank is 1, correctly identifying TUBB2A as the drug target for PCTL. Two other tubulins are the next two most likely candidates. Altogether, 21 proteins are found statistically significant (at $p<0.01$) drug target candidates, of which five tubulin proteins (Supplementary Table 1). Similarly, 5FU dataset gave 15 significant protein candidates, with the known target TYMS being on the $4^{th}$ position. In MTX treatment, the target DHFR ranks $1^{st}$ among seven significant proteins. The TDX dataset features 10 significant proteins, with the main target TYMS on the $1^{st}$ position and the secondary target DHFR on the $3^{d}$ position. DOXO treatment yielded the smallest number of significant proteins (four), which was caused by high cell-type specificity of the drug's action (DOXO primarily acts through DNA intercalation). Intermediate results illustrating the action of "*in silico* purification" for 5FU, TDX, MTX and PCTL are given in Table 1.

**Action mechanism.** Analysis of the interaction networks encompassing protein candidates significantly regulated at $p \leq 0.05$ (32 proteins for 5FU, 9 for DOXO, 13 for MTX, 34 for PCTL and 20 for TDX) highlighted the drug action mechanism. For 5-FU, the main cluster involves ribosome (Figure 2c), the proteins of which were found specifically downregulated. This observation is in line with the previous finding that ribosome suppression is a significant element of 5-FU action.[14,17] Networks for other drugs are shown in Supplementary Figure 1. As would be expected for non-random clusters, a great majority of proteins in clusters with $\geq 3$ molecules have same-sign regulation (up or down). For the largest clusters, the same regulation have 13 out of 15 proteins (5-FU, down); 9 out of 11 (PCTL, up), and 9 out of 10 proteins (TDX, up). According to gene set enrichment analysis, cytosolic large ribosomal subunit (GO cellular component) proteins are overrepresented in 5-FU network (p=1.8E-4); protein polymerization (GO biological process; p=4E-4) in PCTL; and pyrimidine metabolism (KEGG; p=6E-5) in TDX. Thus, drug action mechanisms become apparent from such analysis.

**Death/survival mechanism**. Two-dimensional principal component analysis of 100 most abundant proteins (average of the abundances in three cells lines), with data grouped according to treatments, revealed the evolution of the cellular proteomes, and thus highlighted the death/survival pathways. Figure 1g) demonstrates that the death/survival pathways in MTX and TDX treatments partially overlap, while those in 5-FU treatment are different from these two drugs, despite the fact that both 5-FU and TDX target TYMS.

**Novel targets**. The top protein candidates are likely to be strongly implicated in the mechanism of drug action, and may even represent novel targets for the applied drugs. For instance, overexpression of stratifin (isoform 1 of 14-3-3σ protein), which is found upregulated with a rank 2 in 5-FU treatment, has been suggested to predict resistance to 5-FU therapy in colorectal carcinoma patients.[18] ASNS (glutamine-hydrolyzing asparagine synthetase isoform b) is found downregulated on the 4th position in MTX treatment, which is consistent with this enzyme's action being opposite to that of MTX.[19]

Since starvation/senescence was included in the experiment, the proteomics dataset of this "natural environment" cell suppression could be processed in the same way as other datasets for drug-treated proteomes. In total, 91 proteins were found with p<0.01, much more than in any other treatment, which is consistent with many molecules being targeted by the toxic waste products accumulated in the media. Among the top proteins, there are many known and prospective targets for anticancer therapy. For instance, thioredoxin (rank 3; downregulated) has been identified as a molecule of significant interest to chemotherapy[20], while 3-hydroxyisobutyryl-CoA hydrolase (HIBCH) (upregulated; rank 6) is one of the targets for Quercetin, a molecule found in grape juice, which is being tested as a treatment of prostate cancer by diet (http://clinicaltrials.gov/ct2/show/NCT01912820). The most striking upregulation (Figure 2c) is shown by inter alpha-trypsin inhibitor heavy chain 2 (ITIH2, rank 1). ITIH2 has earlier been suggested to act as a tumor-suppressing protein.[21] We hypothesize that ITIH2 plays an important role in cancer cell survival at adverse conditions and thus represents a potential drug target.

*Experiment II*. In a scaled-down experiment performed for the sake of verification, RKO (colon cancer) and A375 (melanoma) cells were treated with DOXO, 5-FU, camptothecin (CAMP) and PCTL. After data processing, in the 5-FU dataset, seven proteins were found with $p<0.01$, of which TYMS was on the 4$^{th}$ position ($p=7.7 \cdot 10^{-4}$). In the CAMP treatment, out of nine significant proteins, TOP1 (target for CAMP) was on the 9$^{th}$ position with $p=1.1 \cdot 10^{-3}$. For PCTL, the target ß-tubulin (TUBB8) was on the 7$^{th}$ position ($p=1.3 \cdot 10^{-3}$) out of 11 significant proteins (Supplementary Table 2). Therefore, even in this limited experiment with only two cell lines, the drug targets were confined to a small number (on average, nine) protein candidates.

*Experiment III*. To determine the minimal experiment that can provide meaningful data, we have treated HCT116 cells with 5-FU for 24 h in a triplicate and analyzed the extracted proteins against a triplicate control grown for the same time. The experiment was repeated twice, one time using label-free quantification (3,800 proteins quantified), and another time – with TMT-10 quantification (4,800 proteins). In the label-free dataset (median coefficient of variation of protein abundances among the biological replicates was CV ≈ 9%), the target TYMS was on the $28^{th}$ position in terms of absolute regulation, while in the TMT-10 dataset (CV ≈ 5%) – on the $11^{th}$ position. Out of the 11 common proteins found among the top 30 proteins in both datasets, TYMS was on the $8^{th}$ position. Therefore, meaningful results (≤10 potential candidates) can be obtained even from a single biological comparison. The key is obtaining sufficient statistical power, for which a low CV and an adequate number of biological replicates are required.

**Drug target behavior is exceptional.** In order to investigate the question whether drug targets behavior is normal in terms of protein regulation, we built and analyzed predictive models for protein expression levels. The models predicted the regulation of each protein in each treatment based on the regulations of "best friends" of this protein in the same treatment. For this purpose, for each protein and each treatment we identified five proteins whose regulation in all *other* treatments and controls correlated most with that of the protein of interest. Linear regression of the regulations of these five proteins in other treatments provided a model for predicting the regulation of the protein of interest in a given treatment. Figure 2e shows that, as a rule, these predictions strongly (≥3σ) underestimated the experimentally observed regulations for the treatments where the protein was the drug target. Therefore, the targets' regulations were not only strong, but exceptionally strong when a drug targeting them was applied.

This observation prompted us to use the measure of protein exceptional behavior as an independent criterion in *in silico* purification. To quantify this parameter, we measured how exclusion or inclusion of a specific drug treatment affected the correlation of a particular protein expression profile with all other quantified proteins (Supplementary Figure 2). Addition of this criterion to the ranking provided by regulation further narrowed the list of protein target candidates (Supplementary Table 3). For instance, at $p \leq 0.05$ significance, only one protein (TYMS) was identified as a target candidate for 5-FU, as were five proteins for PCTL, including four beta-tubulins. For TDX, two significant proteins were found, including TYMS with rank 1, and for MTX – also two proteins, with DHFR on the $1^{st}$ position. For senescence/starvation, the list of candidates has shortened from 91 to seven, with a known oncogene CCNDBP1 on the $1^{st}$ position and ITIH2 – on the $2^{nd}$.

**Abundant protein behavior reflects that of drug-specific proteins.** The death/survival pathways were determined by mapping the abundance changes in most abundant proteins using supervised principal component analysis (the OPLS-DA[22] method). In order to investigate how the changes in the "top proteome" reflected the behavior of drug-specific proteins, 100 proteins (set *A*) were selected from experiment I with the highest reference abundance[15] (geometric mean of integrated ion current of all unique peptides for all cell types and treatments). In parallel, 100 most drug-specific proteins (set *S*) were selected with the smallest product of p-values calculated for each protein in each drug treatment. The *S*-set encompassed all revealed primary as well as most secondary drug target candidates. The *A*-sets and *S*-sets did not overlap; moreover, on average the *S*-proteins were more than one order of magnitude less abundant than the *A*-proteins. Yet the corresponding OPLS-DA plots were surprisingly similar. For each of the four components that were fitted by the OPLS-DA model, the correlation between the respective components of the *A*- and *S*-models were greater than 0.8 ($R^2$ ranged from 0.70 to 0.87; Figure 3).

The first (most significant) component of the S-model separates DOXO and PCTL on one side from 5-FU, MTX and TDX on the other, with ATP synthase ATP5A1 as the most typical DOXO representative and RNA binding motif protein RBM28 as the 5-FU champion (Supplementary Figure 3 a-b). Among the DOXO/PCTL-specific proteins, there is also a group of tubulins. TYMS is found among the eight most 5-FU/TDX/MTX-specific proteins. The second component (Supplementary Figure 3 c-d) separates 5-FU and TDX/MTX treatments. Here, up-regulation of eukaryotic translation initiation factor EIF4B and eukaryotic translation elongation factor EEF1B2, as well as down-regulation of ribosomal protein RPL23A are most specific for 5-FU, while up-regulation of DHFR, TK1, CDK1 and PRIM1 are specific for TDX/MTX. Interestingly, TYMS is found in the middle group in this component. The component 3 (Supplementary Figure 3 e-f) separates PCTL from DOXO. Up-regulated tubulins and down-regulated BPTF are specific for PCTL, while DOXO-related proteins are CDK2 and SEC14L2. The component 4 (Supplementary Figure 3 a-f) differentiates TDX (the specific group of proteins includes CDK2, PRIM1, TK1 and TYMS) from MTX (SYNJ2, SEC14L2 and DHFR).

**Discussion**

Just a few years ago, cross-comparison of three cells lines at the baseline to the depth of 5,000 proteins has been reported for the first time.[23] Rapid recent progress in proteomics instrumentation and software have led to a marked decrease in the duration of a typical proteomics experiment, enabling analysis of ≥5,000 proteins in the time frame of ≤2 h.[16] This opened a previously unexplored opportunity to apply cellular proteomics to dozens,[8] and in perspective – hundreds and thousands of proteomes,[7] enabling cross-comparison between different cell lines grown at different conditions. This development unlocked the analytical power of the proteome cross-comparison, which created a basis for the current study.

The panel of tested drugs encompasses such diverse mechanisms as DNA and/or RNA synthesis inhibitors (5-FU and TDX), antifolate agents (MTX), tubulin-active antimitotic agents (PCTL), and TOP1 inhibitors (CAMP).[3] The similarity in the formally assigned target or mechanism was not, however, a determinant for the similarity in the death/survival pathway. Indeed, MTX and TDX have formally different mechanisms of actions, but the proteomes of dying/surviving cells were rather similar, while being significantly different from the proteome of cells that underwent 5-FU treatment, which targets the same protein (TYMS) as TDX.

The results of the current study are supportive of the hypothesis that the protein drug target exhibits exceptional regulation compared to other proteins that change their abundance in a way harmonic with the abundances of other co-regulated proteins. It is worth investigating how general this feature is, on a much larger panel of drugs, and with a broader panel of cell lines. The method, if proven general, can significantly shorten drug target identification, which is one of the major bottlenecks in the drug discovery procedure. Its findings need however be verified by orthogonal techniques, such as binding assays. Even with this limitation, high-content proteomics has a chance of becoming an important, perhaps even irreplaceable, tool in drug target discovery.

Within the current experiments, it was not possible to differentiate between the death and survival pathways. A separate study will be needed, where the dying cells are analyzed separately from the surviving cells, preferably in dynamics. Nonetheless, the finding that the OPLS-DA analysis of 100 most abundant proteins mimicked that of 100 most regulated, drug-specific proteins strongly suggests that the mechanisms of drug action and the death/survival pathways are intimately linked. The old paradigm that different triggers of cellular death lead to a generic apoptotic pathway is thus thoroughly rejected. In fact, the opposite seems to be true – each of the tested drug imprinted a unique signature that was easily discernable even on a limited set of household proteins. Thus, "shallow" proteomics analysis monitoring the phenotype evolution of the top cellular proteome and taking as little as 30 min could thus be used in large-scale screening of drug action mechanisms.

**Acknowledgements**

This work was supported by the Knut and Alice Wallenberg Foundation (RZ) and CancerFonden (NV). Liban Abakar is acknowledged for helping with Experiment 2, and Alexander Manoilov – with Experiment 3. Steve Gygi is gratefully acknowledged for providing TMT-10 data in Experiment 3.

**Supporting Information Available**: Mass-spectra (Thermo raw files), extracted peaklists (MFG-files) and database search results (Mascot DAT files) are available on ProteomeXchange. Details of the mass spectrometric analyses used to generate the data, as well as Excel files containing all information presented here.

**FIGURE LEGENDS**

**Figure 1**. General workflow of the proteomics based method of drug target identification.

(a) a panel of cell lines is treated by a panel of drugs and controls, in biological triplicates; (b) LC-MS/MS based proteomics identifies and quantifies ca. 5,000 proteins; (c) for each protein, cell line and treatment, regulation Reg and specificity Spec are calculated; (d) for each treatment, final protein ranks based on Reg and Spec and the p-values are calculated, protein list is sorted in ascending order of p-values; (e) top significant proteins represent the most likely drug targets; (f) top *n* statistically significant proteins are mapped on protein networks to identify the drug target mechanism; (g) most abundant part of the proteomes is mapped using principal component analysis to reveal the cell death/survival pathways.

**Figure 2.** Protein drug targets behavior in the cancer cell lines under the treatment with selected drugs.

Regulation of proteins upon different treatments for three cell lines: (a) TUBB2A, (b) DHFR and (c) ITIH2. (d) Mapping of significant proteins on known networks of protein-protein interactions reveals the mode of 5-FU action via ribosome suppression together with TYMS inhibition. (e) Distributions of prediction errors of protein regulations (regulation – prediction) reveal exceptional behavior of drug targets (3σ area is shaded).

**Figure 3.** OPLS-DA score plots for discriminative models based on two sets of proteins: top 100 most abundant proteins (set A) and top 100 most drug-specific proteins (set S).

(a) OPLS-DA score plot (components 1-2) for set *A*; (b) OPLS-DA score plot (components 1-2) for set *S*; (c-f) corresponding scatter plots (components 1-4) comparing OPLS-DA scores of the A-set and S-set.

**TABLES**

**Table 1.** Ranks for Specificity/Regulation for known protein targets (in parentheses) of four drugs in different cell lines.

| Cell line | 5-FU (TYMS) | MTX (DHFR) | PCTL (TUBB2A) | TDX (TYMS) |
|---|---|---|---|---|
| Melanoma | 44/70 | 1/1 | 38/12 | 18/3 |
| Lung | 11/7 | 2/2 | 7/12 | 65/11 |
| Colon | 136/813 | 8/9 | 55/30 | 11/8 |
| Final | 4 | 1 | 1 | 1 |

**SUPPLEMENTARY MATERIALS**

**Supplementary Table 1**. Label-free proteomics quantification data and protein drug target identification results based on Regulation and Specificity for Experiment 1. Cell lines HCT116, A375, and H1299 were treated with 5-FU, TDX, MTX, PCTL and DOXO, as well as DMSO (control), or left in unchanged media for senescence/starvation (SEN).

**Supplementary Table 2**. Label-free proteomics quantification data and protein drug target identification results based on Regulation and Specificity for Experiment 2. Cell lines RKO and A375 treated with DOXO, 5-FU, CAMP and PCTL, as well as DMSO (control), or left in unchanged media for senescence/starvation (SEN).

**Supplementary Table 3**. Drug target identification results based on drug target exceptional behavior for Experiment 1. Cell lines HCT116, A375, H1299 treated with 5-FU, TDX, MTX, PCTL and DOXO, as well as DMSO (control), or left in unchanged media for senescence/starvation (SEN).

**Supplementary Table 4.** Lists of top 100 most abundant proteins (set A) and top 100 most drug-specific proteins (set S).

**Supplementary Figure 1**. Protein-protein interaction networks for significant drug target protein candidates (rank-product p≤0.05) for 5-FU, TDX, MTX, PCTL and DOXO.

**Supplementary Figure 2**. Correlations of specific proteins profiles (TYMS and TUBB2A) with all other protein profiles in all treatments (x-axis) and excluding treatments with specific drugs (5-FU and PCTL) (y-axis).

**Supplementary Figure 3**. OPLS-DA loading and score plots for discriminative model for 5 drugs (PCTL, DOXO, 5-FU, MTX and TDX) based on label-free quantitative data of 100 most drug-specific proteins: (a) scores and (b) loadings for components 1-4; (c) scores and (d) loadings for components 2-4; (e) scores and (f) loadings for components 3-4.

**ONLINE METHODS**

**Cell culture and drug treatments.** HCT116 and RKO (colon carcinoma), H1299 (lung cancer), and A375 (melanoma) cell lines were kindly provided by colleagues from Karolinska Institutet. The cells were cultured at 37 °C with 5% $CO_2$ in high-glucose Dulbecco's Modified Eagle's Medium (DMEM, Thermo Fisher Scientific) supplemented with 10% fetal bovine serum (Gibco) and 1% penicillin/streptomycin (Gibco). The cells were treated for 24-96 h with six different drugs: 5-fluorouracil (5FU), raltitraxed or tomudex (TDX), metotrexate (MTX) (all - Sigma), as well as doxorubicin (DOXO), paclitaxel (PCTL), and camptothecin (CAMP) (all - Eurasia Drugs, China). Each type of cell was treated with a concentration causing death of 15-50% of cells after 48 h of treatment. In Experiment 1, the following was used:

| CELL LINE | 5FU | TDX | DOXO | PCTX | MTX | CONTROLS |
|---|---|---|---|---|---|---|
| HCT 116 | 50 μM | 100 nM | 5 μM | 100 nM | 5 μM | 0h, 48h, SEN |
| A375 | 10 μM | 50 nM | 100 nM | 50 nM | 100 nM | 0h, 48h, SEN |
| H1299 | 50 μM | 10 μM | 15 μM | 100 nM | 5 μM | 0h, 48h, SEN |

In Experiment 2, the concentrations were 10 μM for 5-FU, 20 μM for TDX, 0.5 μM for PCTL, 30 nM for DOXO, and 15 μM for CAMP.

All drugs were dissolved in 0.01% dimethyl sulfoxide (DMSO). As a negative control, cells were treated with 0.01% DMSO. The medium supplemented with the drug was replaced each 24 h by fresh medium, except for starved/senescent cells that were left in the same medium for 10 days.

*Protein extraction and digestion.* The collected cells were suspended in lysis buffer (1 mln cells in 100 µL buffer). The buffer was prepared by dissolving 1 mg ProteaseMax (Promega) in 900 µL ammonium bicarbonate (50 mM) and 100 µL acetonitrile (ACN). ProteaseMax is a surfactant which not only solubilizes the proteins but enhances subsequent tryptic digestion of proteins as well. The samples were vortexed for 5 min and then heated in shaking thermomixer (Eppendorf) at 50 °C for 30 min at 1400 rpm, followed by sonication for 30 min. The total protein concentration was measured using the BCA protein assay kit (Pierce) in accordance with the manufacturer's protocol. The extracted proteins were reduced with 5.5 mM dithiothreitol (DTT), alkylated with 15 mM indole-3-acetic acid (IAA), and digested with 1.2 µg modified sequencing grade trypsin (Promega) dissolved in 50 mM ammonium bicarbonate. After 14 h of tryptic digestion, the reaction was stopped with acetic acid to a final concentration of 5% and then heated to 56 °C for 30 min at 500 rpm, followed by centrifugation for 7 min at 14,000 rpm at room temperature. The samples were pre-cleaned in a C18 column Zip-tip (Millipore), and the flow-through was dried in a SpeedVac centrifugal evaporator. The dried peptides were dissolved in water containing 1% formic acid (Fluka) for LC-MS/MS analysis. The above described digestion protocol was performed using the Mass Prep Station Robotic Protein Handling System (Waters, Manchester, UK).

*LC-MS/MS experiment.* For each sample, 5 µg of peptides were analyzed using Orbitrap Q Exactive (Experiments 1, 3/label-free), Orbitrap Velos (Experiment 2) or Orbitrap Fusion (Experiment 3/TMT-10) mass spectrometers coupled to a nEasy HPLC (all - Thermo Fisher Scientific). Chromatographic separation of peptides was achieved using a 50 cm Easy nanoflow column (Thermo; Experiments 1 and 3) or a 75 µm ID fused silica column packed in-house (Experiment 2) to the length of 8 cm with a slurry of reverse-phase, fully end-capped Reprosil-Pur C18-AQ 3 μm resin in methanol. The peptides were loaded onto the column at a flow rate of 1000 nL/min, and then eluted at a 300 nL/min flow rate for 180-210 min at a linear or biphasic gradient from 4% to 35% ACN in 0.1% formic acid. Electrospray ionization of the peptides was at 1.5 kV. The MS and MS/MS data was acquired in the Orbitrap mass analyzer in a data-dependent acquisition mode. The survey MS spectrum was acquired at the resolution of 60,000 in the range of m/z 200 − 2000. MS/MS data were obtained with a higher-energy collisional dissociation (HCD) for ions with charge z≥2 at a resolution of 7,500 (Orbitrap Velos) or 15,000 (Q Exactive and Fusion).

*Data processing.* The raw files were converted to Mascot Generic Format (mgf) using in-house written Raw2mgf program. All mgf files were merged to create a common mgf file using in-house written Cluster program, which merged individual MS/MS spectra sharing more than 12 out of 20 most abundant peaks. The clustered mgf files were searched by the MS/MS search engine Mascot (version 2.3.0, Matrix Science, UK) to identify peptides and proteins. The mass tolerance was 10 ppm for precursor ions and 20 mDa for fragment ions, using carbamidomethyl (C) as a fixed modification, oxidation (M) as a variable modification, and up to two missed tryptic cleavages. The IPI human database (version 3.68; 91,521 human protein sequences) was searched, with reversed protein sequences concatenated as a decoy for determining the false discovery rate (FDR).

Quantitative information was extracted using in-house developed label-free software Quanti v.2.5.3.1.[15] Only reliably identified (FDR<0.01), unmodified peptides with unique sequences were considered and only proteins discovered with at least two such peptides were quantified. For each protein, one database identifier (ID) was selected, covering all the peptide sequences identified for this specific protein. If two proteins belonging to different protein groups had a partial sequence overlap, then all the peptides belonging to this overlap were ignored. The results were reported as a set of relative protein abundances *A* scaled such that the geometric mean of the abundance of each protein over all samples was 1.0.

*Scoring system.* For combining the data from replicate analysis, "medians of ratios" are used instead of "ratios of medians", as has previously been suggested.[24] If relative protein abundance of *i*-th quantified protein in *c*-th cell line under *j*-th treatment is denoted as $A_{i,j}^c$, then regulation Reg is calculated as:

$$\boldsymbol{Reg}_{i,j}^c = Median\left(\left|\log \frac{A_{i,j}^c}{A_{i,0}^c}\right|\right), \quad (1)$$

and specificity *Spec* is defined as:

$$\boldsymbol{Spec}_{i,j}^c = Median_{k \neq j}\left(\left|\log \frac{A_{i,j}^c}{A_{i,k}^c}\right|\right), \quad (2)$$

where j=0 corresponds to untreated cells for *Reg* calculation, and j≠k for *Spec* calculations.

For each cell line and treatment, the proteins were sorted by *Reg* and *Spec*, making sure that the direction of *Reg* and *Spec* in top proteins were the same (otherwise the proteins were forced to the bottom of the list). Then the *Reg* and *Spec* ranks were summed for all cell lines, and the proteins were sorted in ascending order of the summed rank. Top proteins represented the most likely drug target candidates.

*Exceptional behavior measure.* For each *I*-th protein and each *J*-th drug treatment, two vectors were calculated:

$$C_i^{I,*} = Corr(\mathbf{Reg}_{i,j}^C, \mathbf{Reg}_{I,j}^C),$$

$$C_i^{I,J} = Corr(\mathbf{Reg}_{i,j\neq J}^C, \mathbf{Reg}_{I,j\neq J}^C),$$

where $C_i^{I,*}$ are the Pearson's correlation coefficients of expression profiles over all treatments of *i*-th and *I*-th proteins, while $C_i^{I,J}$ are correlation coefficients of the expression profiles of *i*-th and *I*-th proteins <u>excluding</u> treatment J. Then, the linear model $C_i^{I,*} \sim C_i^{I,J}$ was created and the coefficient of determination of the model was used to calculate the measure of exceptional behavior $E^{I,J}$ of *I*-th protein under *J*-th treatment:

$$E^{I,J} = \frac{1}{\overline{R^2}_{I,J}}$$

*p-value calculation.* In estimation of the p-value of a protein with a certain rank, we used the rank product method, which has previously been found to be robust and tolerant to missing values in detection differentially regulated genes in replicated experiments.[25] The method has also been successfully applied to proteomics datasets for detection of significantly regulated proteins.[26] In adaptation of the method by Schwämmle et al., we treated **Reg** and **Spec** ranks as independent variables, and their values for different cell lines as well as at different incubation times were considered as independent replicate measurements. The rank product was considered to have a gamma distribution under null hypothesis, from which we calculated the p-values for the set of ranks of every protein. Adjusted p-values were calculated using standard Bonferroni correction, using the total number of proteins as a multiplication factor.

# REFERENCES

1. Lee, J. A., Uhlik, M. T., Moxham, C. M., Tomandl, D. & Sall, D. J. Modern phenotypic drug discovery is a viable, neoclassic pharma strategy. *J. Med. Chem.* **55,** 4527–38 (2012).
2. Holbeck, S. L., Collins, J. M. & Doroshow, J. H. Analysis of Food and Drug Administration-approved anticancer agents in the NCI60 panel of human tumor cell lines. *Mol. Cancer Ther.* **9,** 1451–60 (2010).
3. Reinhold, W. C. *et al.* CellMiner: a web-based suite of genomic and pharmacologic tools to explore transcript and drug patterns in the NCI-60 cell line set. *Cancer Res.* **72,** 3499–511 (2012).
4. Schirle, M., Bantscheff, M. & Kuster, B. Mass spectrometry-based proteomics in preclinical drug discovery. *Chem. Biol.* **19,** 72–84 (2012).
5. Zubarev, R. a. The challenge of the proteome dynamic range and its implications for in-depth proteomics. *Proteomics* **13,** 723–6 (2013).
6. Kim, M.-S. *et al.* A draft map of the human proteome. *Nature* **509,** 575–581 (2014).
7. Wilhelm, M. *et al.* Mass-spectrometry-based draft of the human proteome. *Nature* **509,** 582–587 (2014).
8. Gholami, A. M. *et al.* Global proteome analysis of the NCI-60 cell line panel. *Cell Rep.* (2013).
9. Geiger, T., Wehner, A., Schaab, C., Cox, J. & Mann, M. Comparative proteomic analysis of eleven common cell lines reveals ubiquitous but varying expression of most proteins. *Mol. Cell. Proteomics* **11,** M111.014050 (2012).
10. Vogel, C. & Marcotte, E. M. Insights into the regulation of protein abundance from proteomic and transcriptomic analyses. *Nat. Rev. Genet.* **13,** 227–32 (2012).
11. Martinez Molina, D. *et al.* Monitoring drug target engagement in cells and tissues using the cellular thermal shift assay. *Science* **341,** 84–7 (2013).
12. Cohen, a a *et al.* Dynamic proteomics of individual cancer cells in response to a drug. *Science* **322,** 1511–6 (2008).
13. Good, D. M. & Zubarev, R. a. Drug target identification from protein dynamics using quantitative pathway analysis. *J. Proteome Res.* **10,** 2679–83 (2011).
14. Marin-Vicente, C., Lyutvinskiy, Y., Romans Fuertes, P., Zubarev, R. A. & Visa, N. The effects of 5-fluorouracil on the proteome of colon cancer cells. *J. Proteome Res.* **12,** 1969–79 (2013).
15. Lyutvinskiy, Y., Yang, H., Rutishauser, D. & Zubarev, R. In silico instrumental response correction improves precision of label-free proteomics and accuracy of proteomics-based predictive models. *Mol. Cell. Proteomics* 1–26 (2013). doi:10.1074/mcp.O112.023804
16. Pirmoradian, M. *et al.* Rapid and Deep Human Proteome Analysis by Single-dimension Shotgun Proteomics. *Mol. Cell. Proteomics* **12,** 3330–8 (2013).
17. Burger, K. *et al.* Chemotherapeutic drugs inhibit ribosome biogenesis at various levels. *J. Biol. Chem.* **285,** 12416–25 (2010).
18. Perathoner, A. *et al.* 14-3-3sigma expression is an independent prognostic parameter for poor survival in colorectal carcinoma patients. *Clin. Cancer Res.* **11,** 3274–9 (2005).
19. *Small Animal Clinical Pharmacology*. 360 (Elsevier Health Sciences, 2008). at <http://www.google.se/books?id=RpsROVqemk8C&pgis=1>

20. Arnér, E. S. J. & Holmgren, A. The thioredoxin system in cancer. *Semin. Cancer Biol.* **16,** 420–6 (2006).
21. Hamm, A. *et al.* Frequent expression loss of Inter-alpha-trypsin inhibitor heavy chain (ITIH) genes in multiple human solid tumors: a systematic expression analysis. *BMC Cancer* **8,** 25 (2008).
22. Bylesjö, M. *et al.* OPLS discriminant analysis: combining the strengths of PLS-DA and SIMCA classification. *J. Chemom.* **20,** 341–351 (2006).
23. Lundberg, E. *et al.* Defining the transcriptome and proteome in three functionally different human cell lines. *Mol. Syst. Biol.* **6,** 450 (2010).
24. Brody, J. P., Williams, B. a, Wold, B. J. & Quake, S. R. Significance and statistical errors in the analysis of DNA microarray data. *Proc. Natl. Acad. Sci. U. S. A.* **99,** 12975–8 (2002).
25. Breitling, R., Armengaud, P., Amtmann, A. & Herzyk, P. Rank products: a simple, yet powerful, new method to detect differentially regulated genes in replicated microarray experiments. *FEBS Lett.* **573,** 83–92 (2004).
26. Schwämmle, V., León, I. R. & Jensen, O. N. Assessment and Improvement of Statistical Tools for Comparative Proteomics Analysis of Sparse Data Sets with Few Experimental Replicates. *J. Proteome Res.* (2013). doi:10.1021/pr400045u

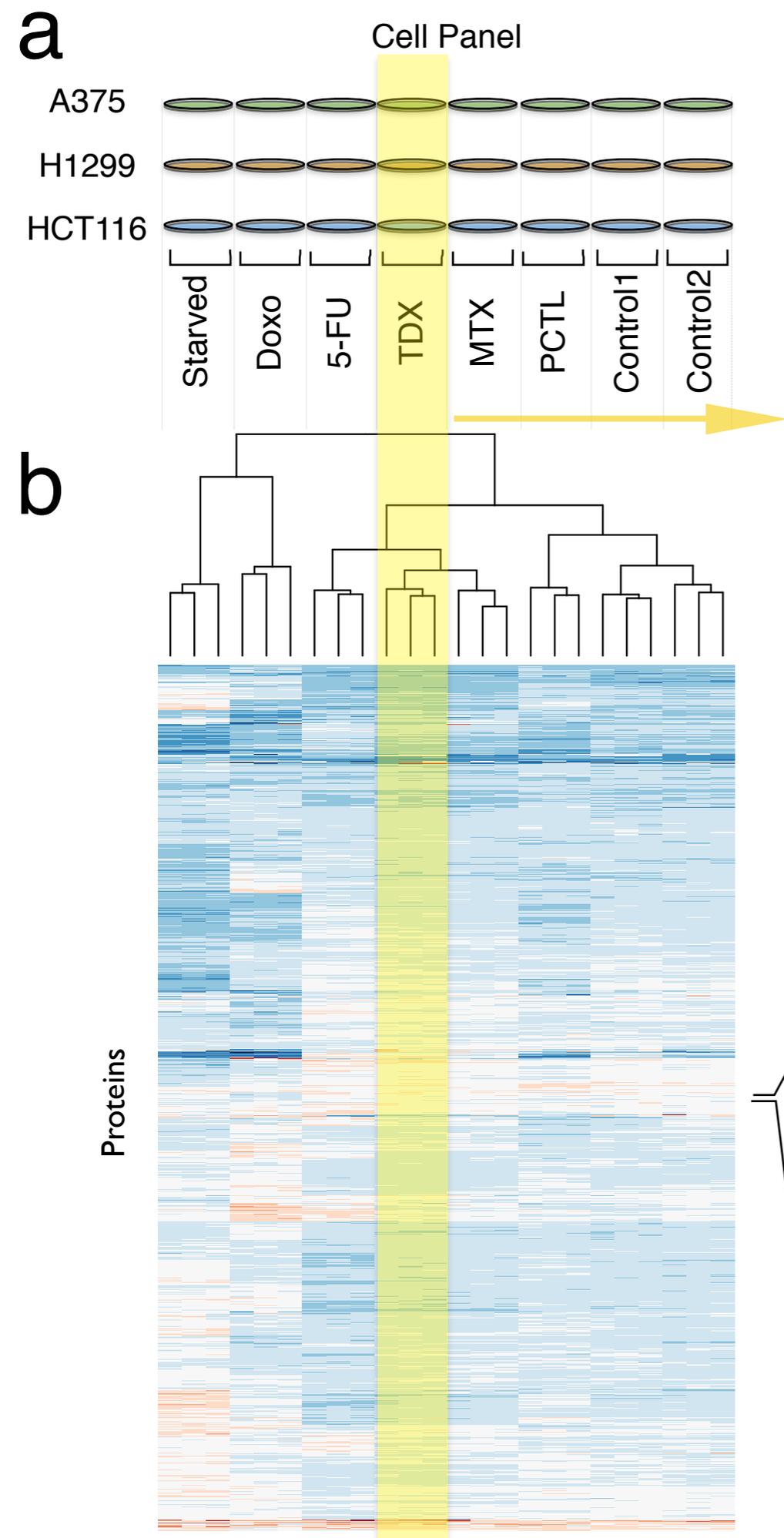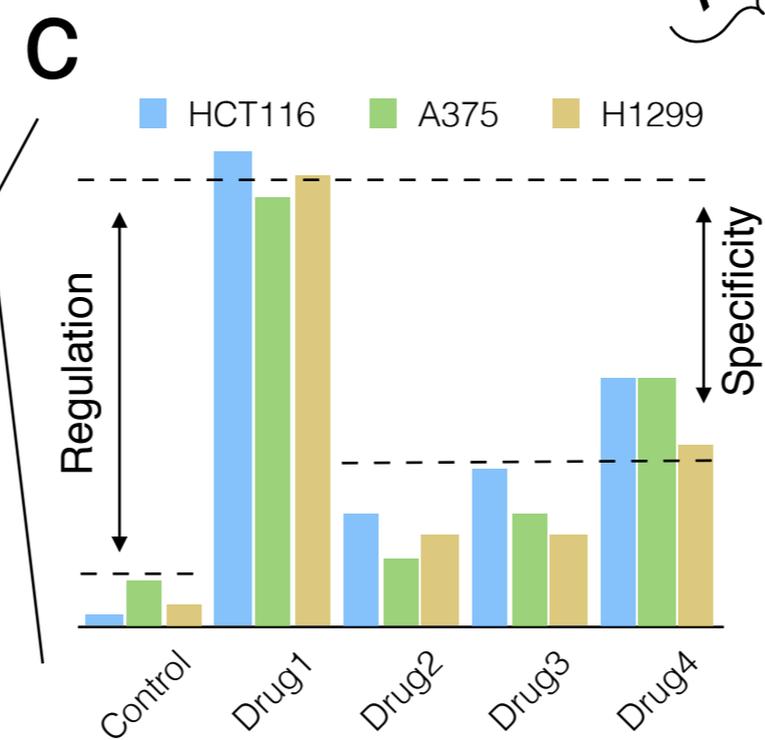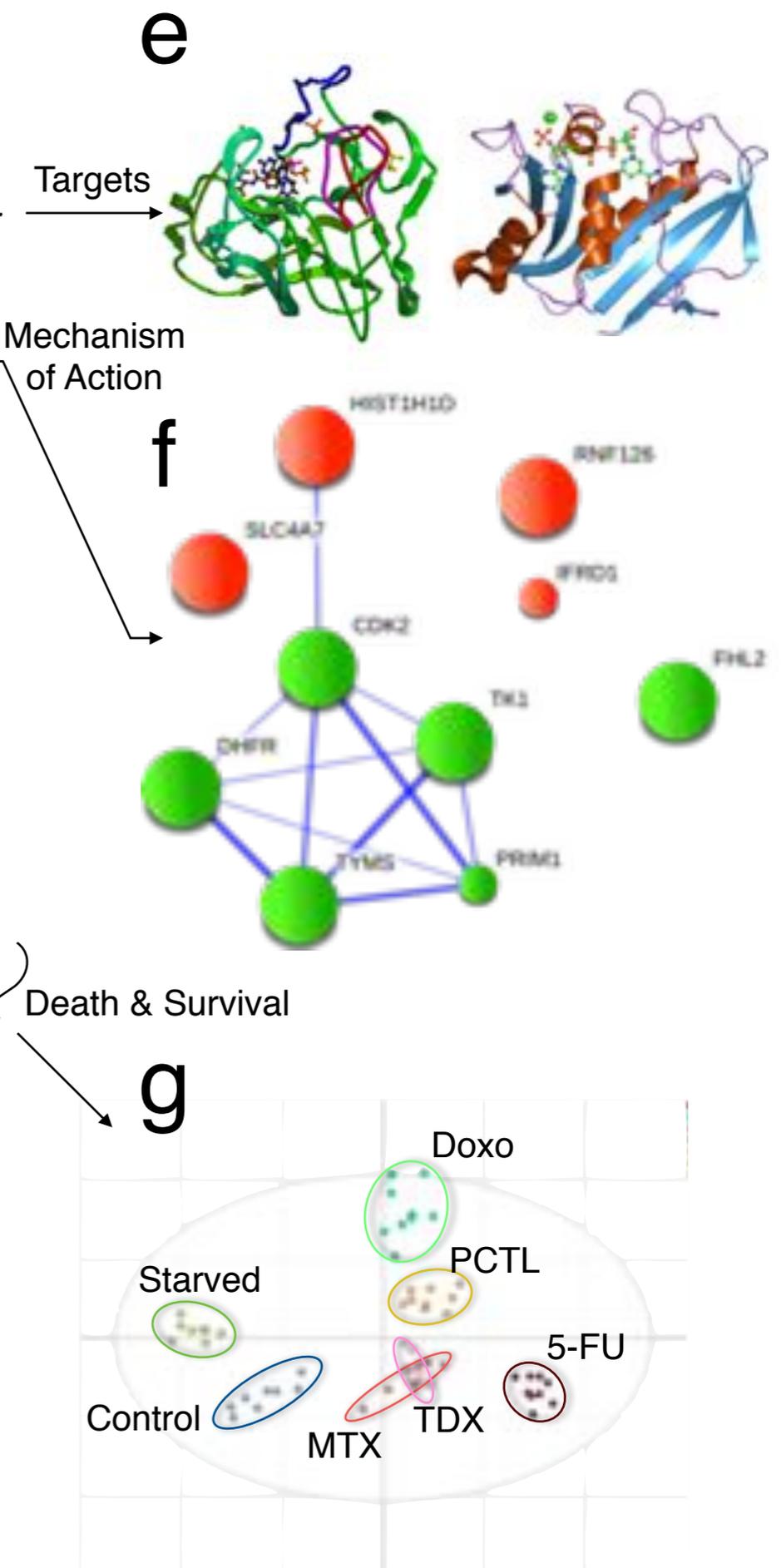

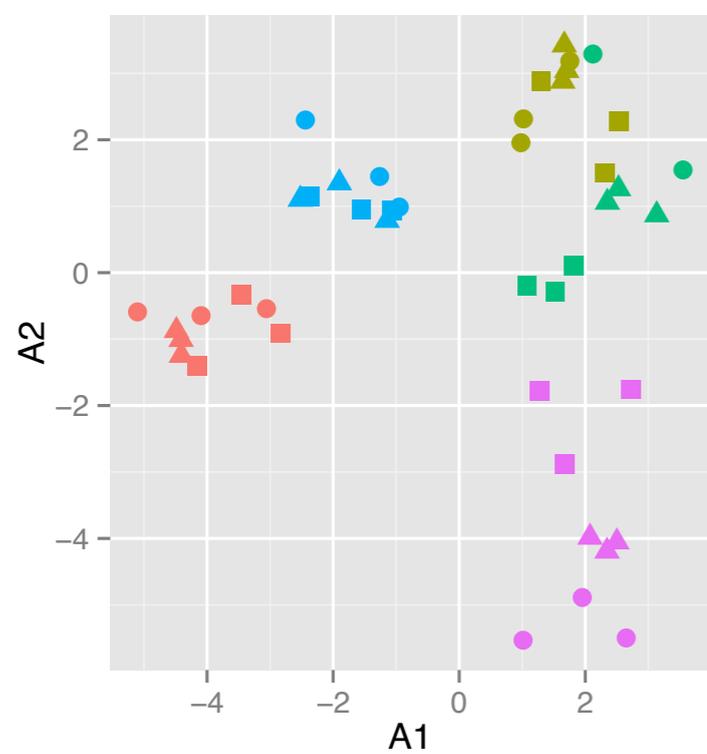
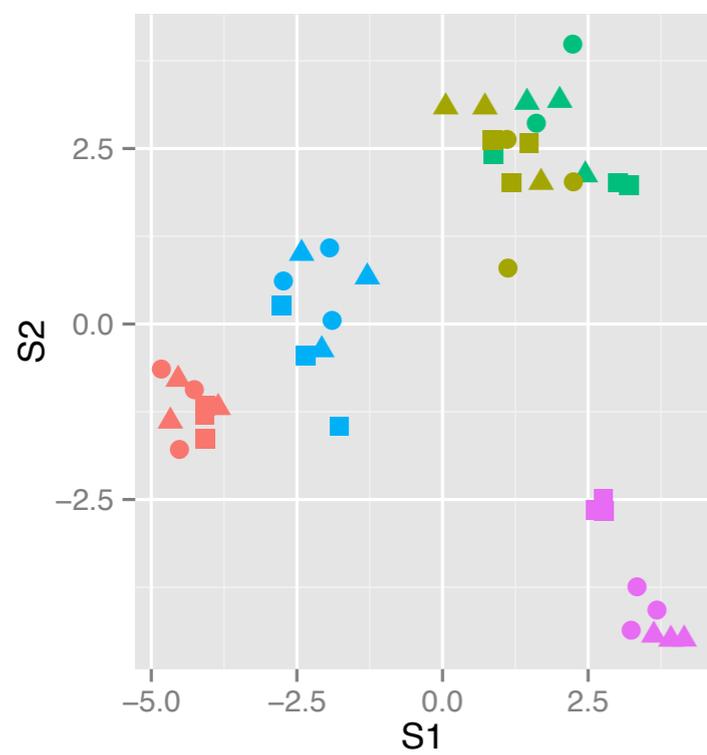
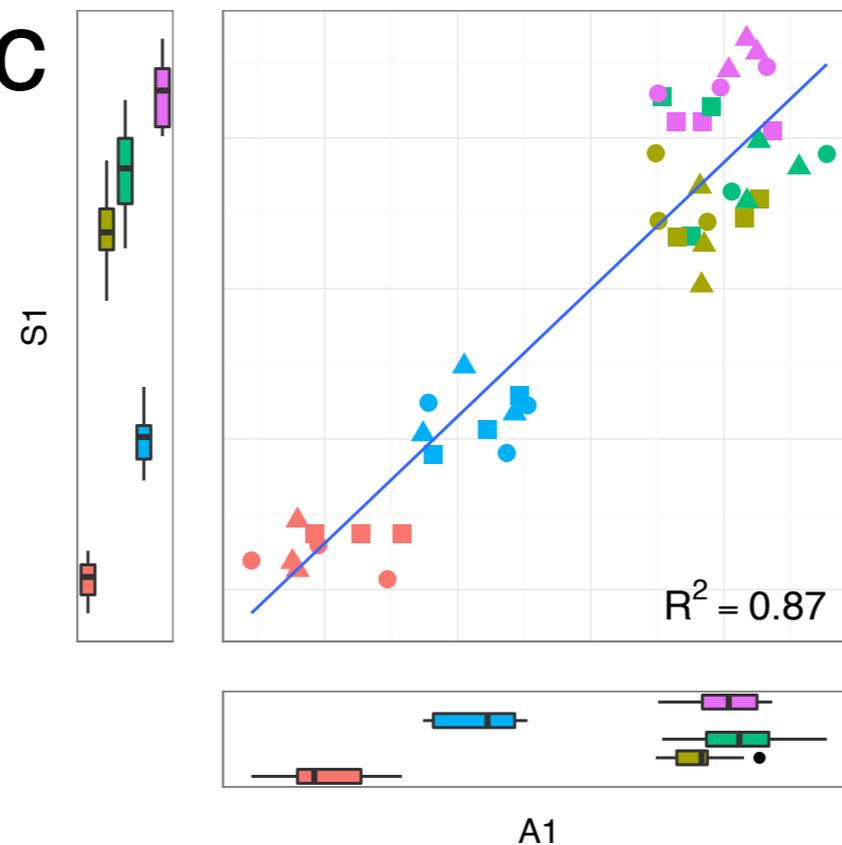
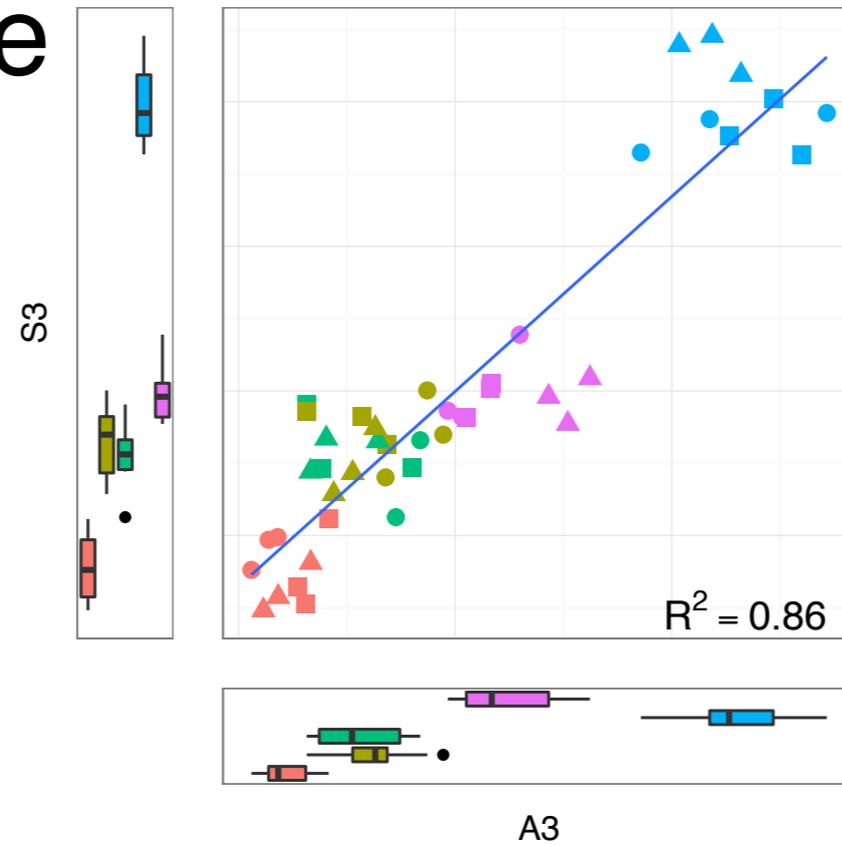
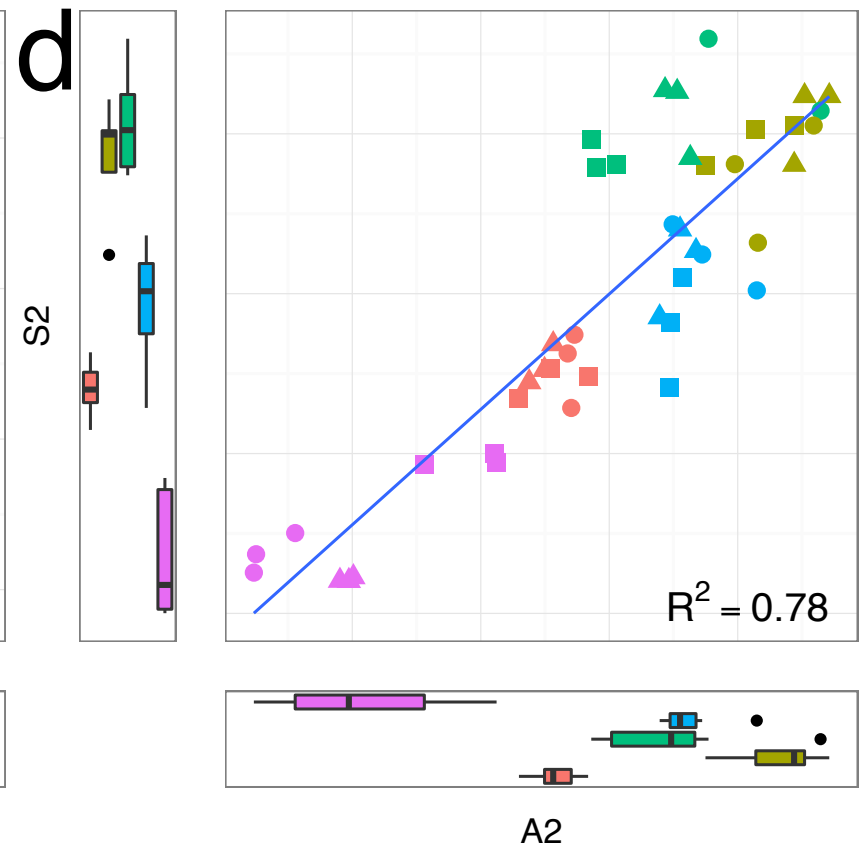
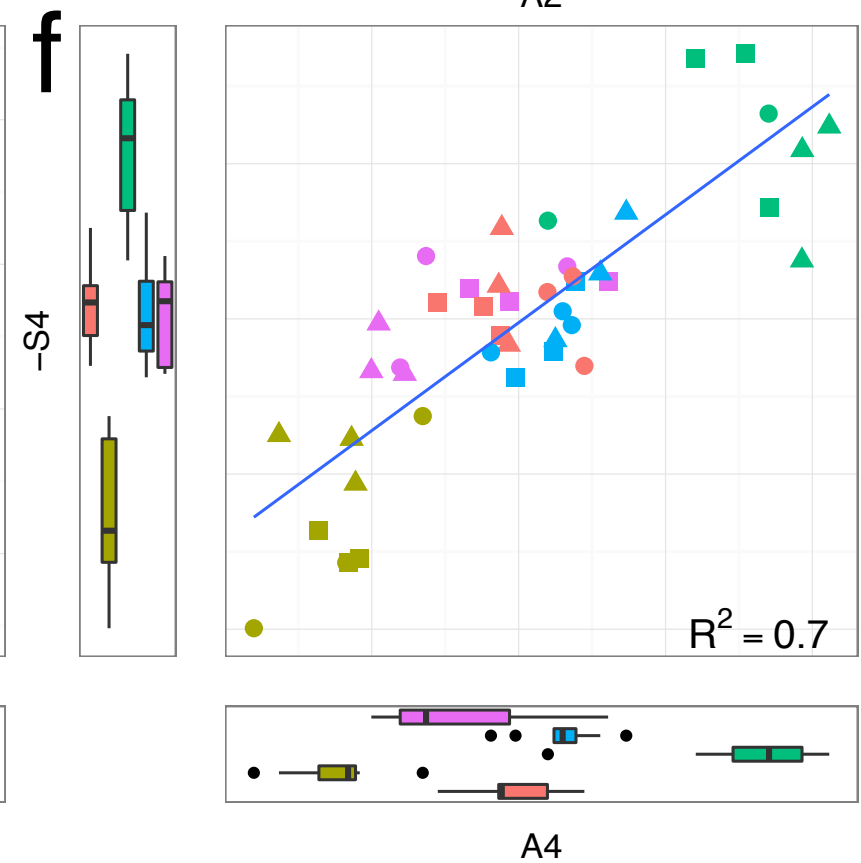